\documentclass[]{AO4ELT}  

\usepackage{microtype}
\usepackage{biblatex}
\usepackage{amsmath,amsfonts,amssymb}
\usepackage{graphicx}
\usepackage{pst-all} 
\usepackage[colorlinks=true, allcolors=blue]{hyperref}
\usepackage{float}

\makeatletter         
\def\@maketitle{
\includegraphics[width = 170mm]{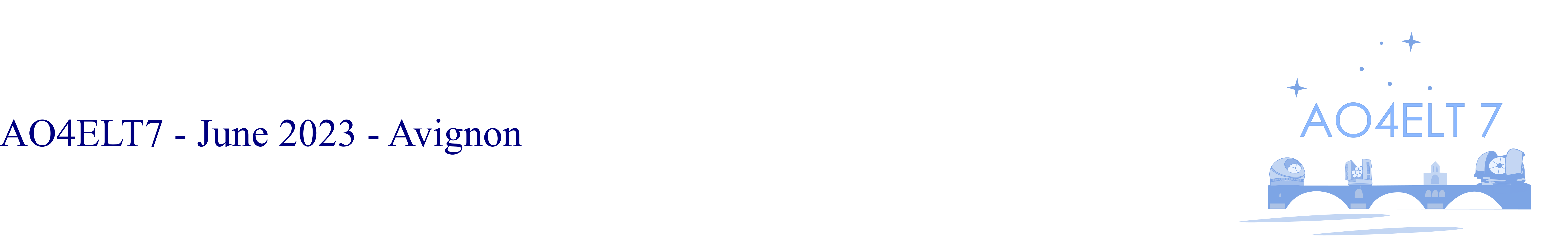}\\[8ex]
\begin{center}
{\Huge \bfseries \sffamily \@title }\\[4ex] 
{\Large  \@author}\\[4ex] 
\@date
\end{center}}

\title{Measurement of isoplanatic angle and  turbulence strength profile from H-alpha images of the Sun}

\author[a,b]{Saraswathi Kalyani Subramanian}
\author[a]{Sridharan Rengaswamy}
\affil[a]{Indian Institute of Astrophysics, 2nd Block Koramanagala, Santhoshapuram, Bengaluru, India}
\affil[b]{Department of Applied Optics and Photonics, University of Calcutta,  87/1, College Street, Kolkata, India}

\authorinfo{Further author information: (Send correspondence to S.S.K.)
\\S.S.K.: E-mail: saraswathi.kalyani@iiap.res.in}

\begin{document} 
\maketitle
\begin{abstract}
Adaptive Optics (AO) systems have become integral for ground-based astronomy. Based on the scientific case, there are various flavours of AO systems. Measuring the turbulence strength profile ($C_N^2(h)$) and other site characteristics is essential before selecting a site or implementing certain types of AO systems. We used an iterative deconvolution procedure on long-exposure H-$\alpha$ images of the Sun to determine the isoplanatic patch size during the daytime. Then, we determined the relationship between turbulence along different directions and also obtained an analytical estimate of the $C_N^2(h)$ profile. 
\end{abstract}

 
\section{INTRODUCTION}
\label{sec:intro}  
Adaptive optics systems are used in ground-based telescopes to correct for the deleterious effects of the Earth’s atmospheric turbulence on the light coming from celestial objects. A classical/Single-Conjugate AO (SCAO) system corrects for the turbulence over a narrow field of view that is typically of the order of a few arc-seconds. For scientific objects that are of larger angular extent (eg: the sunspots on the surface of the Sun, bright solar system planets, the lunar surface,  etc...), it is necessary to resolve the features. AO systems aid this by improving the contrast and resolution that were reduced by the turbulent atmosphere. In such cases, Multi-Conjugate AO (MCAO) systems are used to increase the corrected field of view to 1-2 arcminutes. To achieve this, an MCAO uses multiple deformable mirrors conjugated to different altitudes of the atmosphere.  It is necessary to have knowledge of the turbulence strength profile at the telescope site before implementing such a system.

There are many methods to estimate the turbulence profile at a site, such as SLODAR \cite{2002MNRAS.337..103W}, SCIDAR \cite{1973JOSA...63..270V}, and S-DIMM+ \cite{2010A&A...513A..25S}. In this paper, we propose a new method that can provide an analytical estimate of the turbulence profile using long-exposure images of the Sun. It is an extension of the Parametric Search Method (PSM)\cite{2019SoPh..294....5R}, which uses the long exposure images to estimate the Fried’s parameter ($r_0$) at a given site.  

The paper is organised as follows: Section \ref{sec:psm} recaps the PSM and how we have extended it to measure the isoplanatic patch size. Section \ref{sec:additional_info} describes how the extended PSM can be used to obtain additional information about the site. Section \ref{sec:summary} summarises the results and presents the future scope of our work.

\section{ISOPLANATIC PATCH ANGLE from the PSM}
\label{sec:psm}

The isoplanatic angle is the angular size of the sky region (alternatively, the angular size at the image plane) over which the point-spread function is invariant.  In terms of the perturbed wave-fronts, it can be defined as the angular separation over which there is a significant correlation between the wave-fronts arriving at the telescope's pupil from two different directions. Quantitatively, it is be defined as the angular separation over which the difference between the root mean square wave-font errors of the wave-fronts traversing in those directions is equal to a radian.  

PSM uses long-exposure H-$\alpha$ images of the Sun to estimate $r_0$ \cite{2019SoPh..294....5R, 2021SoPh..296...65U}. Overlapping segments of the image, each deconvolved with a set of long-exposure seeing-limited transfer functions (corresponding to unique values of $r_0$), results in a plot of the ratios of the contrasts of the initial and deconvolved images as a function of $r_0/$Dia. The point of intersection of the two straight lines fitted to the ends of the plot gives an estimate of $r_0$. Applying the process to a large number of images gives the site characteristics (measured in terms of $r_0$). We are interested in knowing if we can exploit the extended nature of our source to extract more information about the site.

\subsection{Extending the PSM}
\label{subsec:extend_psm}

\begin{figure} [H]
   \begin{center}
   \begin{tabular}{c} 
   \includegraphics[height=7cm]{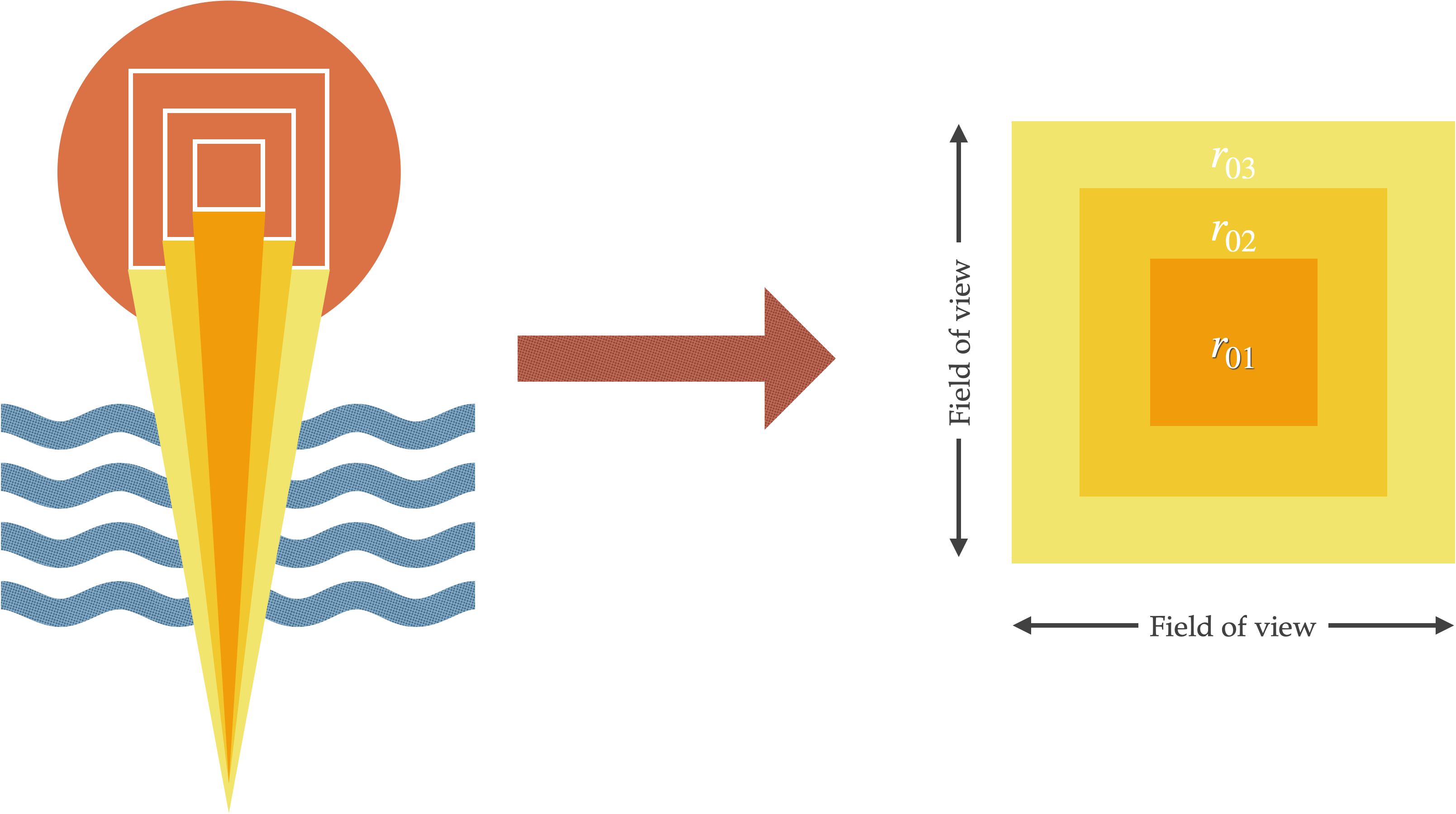}
   \end{tabular}
   \end{center}
   \caption[example] 
   { \label{fig:fov_r0_map} Cartoon representation (not to scale) of process of estimating $r_0$ map from extended PSM. The left portion of the figure shows how choosing different fields of view on the Sun samples different regions of the atmosphere. The right portion shows the $r_0$ map that is built as a function of field of view.}

\end{figure}

An extended source implies that the wave-fronts from it sample a large volume of the atmosphere. So, we extend PSM by increasing the image size used for the deconvolution process in each iteration. By applying PSM to images of different sizes, we obtain a measurement of $r_0$ for each iteration. This gives us $r_0$ as a function of the field of view. Choosing a larger image implies that we are considering wave-fronts coming from a wider angle - hence we are sampling more regions of the atmosphere. This has been shown in Figure \ref{fig:fov_r0_map}. The left portion of the figure shows how choosing different fields of view (marked by three white squares of different sizes) on the surface of the Sun (orange circle) allow us to sample different volumes (yellow and orange triangles) of the atmosphere (blue waves). This allows us to build an $r_0$ map (shown on the right part of Figure \ref{fig:fov_r0_map}) with the colours in the map corresponding to those of the triangles denoting the regions of atmosphere sampled.

\begin{figure} [H]
   \begin{center}
   \begin{tabular}{c} 
   \includegraphics[height=7cm]{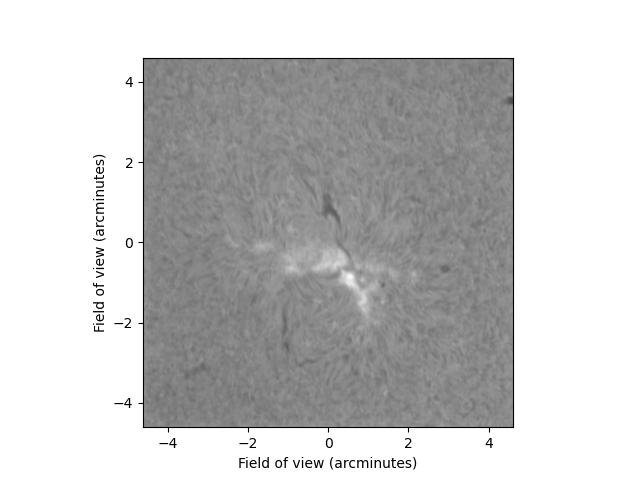}
   \end{tabular}
   \end{center}
   \caption[example] 
   { \label{fig:sol_img} H-$\alpha$ image of the Sun recorded at Merak on 26$^\text{th}$ April, 2018 that has been used for the analysis.}

\end{figure}

We have used data obtained at Merak, Ladakh (on 26$^\text{th}$ April, 2018) with a 20 cm telescope \cite{2018JApA...39...60R}. The images were recorded with a Lyot filter at H-$\alpha$ (656.3 nm, passband of 0.5 \AA) with an exposure time of 700 ms. The sampling was 0.27arc-seconds/pixel and the total image size was around 9.2x9.2 arc-minute$^2$. For the results presented in this paper, we have used 4 non-overlapping regions of one image. The image used in our analysis is shown in Figure \ref{fig:sol_img}.

\begin{figure} [H]
   \begin{center}
   \begin{tabular}{c} 
   \includegraphics[height=7cm]{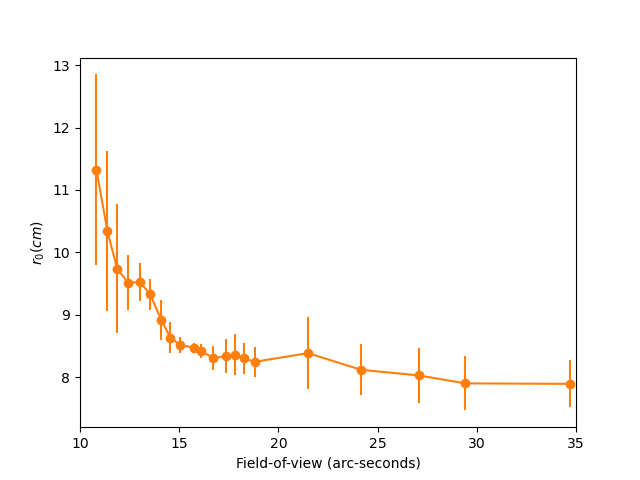}
   \end{tabular}
   \end{center}
   \caption[example] 
   { \label{fig:r0_vs_angle} 
Fried's parameter ($r_0$ in cm) measured as a function of field of view (in arc-seconds) from long exposure H-$\alpha$ images of the Sun using the extended PSM.}
\end{figure}

Figure \ref{fig:r0_vs_angle} shows the measured $r_0$ as a function of size of the field of view  (in arc-seconds). The error bars arise from averaging the measurements by considering different regions of an image for deconvolution. It should be noted that for each value of field of view, the $r_0$ measured using PSM from four non-overlapping regions were averaged. The error bars are the standard deviations of the four values.

From the figure, we can infer that as the angular size of the image used for deconvolution increases, the measured value of $r_0$ decreases. This  can be understood as follows: At a very narrow field of view,  there is a significant overlap between the air columns through which the constituent wave-fronts traverse before reaching the finite-sized telescope pupil. Furthermore, at these narrow angles, the relative contribution of the high-altitude turbulence is significant. As the field of view increases, the degree of overlap between the constituent wave-fronts decreases. The resultant wavefront at the pupil encompasses the turbulent effects over the  frustum of a cone whose volume increases with the field of view. The area covered by the wavefronts at higher layers also increases. This makes larger fields of view more sensitive to ground-layer turbulence which is higher during the daytime. We can see from Figure \ref{fig:r0_vs_angle} that the $r_0$ values begin to plateau beyond a certain field of view - we identify this as the isoplanatic patch size for our long exposure images. Beyond this, we do not see a further degradation of $r_0$. We estimate the isoplanatic patch size to be around 17 arc-seconds from the data we considered. Wang \cite{WANG1975200} has shown that the isoplanatic patch size is dependent on the spatial resolution and that for low resolution seeing-limited (long exposure) imaging with small telescopes, the isoplanatic size could be quite larger than that expected for diffraction-limited imaging. Also, in the case of daytime imaging, the isoplanatic size is expected to be larger than that at night time owing to lower altitude of the seeing layer due to ground heating.



\section{Estimating Turbulence Strength Profile}

\label{sec:additional_info}

\subsection{Angular correlation of wavefront expressed using Zernike Polynomials}
\label{subsec:angular_corr}

The values of $r_0$ measured in section \ref{subsec:extend_psm} are used to generate phase screens described by the Kolmogorov theory of atmospheric turbulence. The phase screens are then decomposed into Zernike polynomials. We obtain a set of coefficients (limited to the 5$^\text{th}$ radial order in our case) for each phase screen. The correlation of a given polynomial coefficient with that from a different phase screen estimates the relation between the wave-fronts propagating through the different angular regions of the atmosphere.

If $\phi_1$ and $\phi_2$ are two phase screens, their Zernike decomposition is

\begin{equation}
\label{eq:zernike}
\phi_1 = \sum_i a_{i1}Z_{i1} \text{   and  }  \phi2 = \sum_i a_{i2}Z_{i2} \, ,
\end{equation}

The correlation of the coefficients is
\begin{equation}
\label{eq:corr}
C_{ii}^{12} = \frac{\langle a_{i1}a_{i2} \rangle}{\sqrt{\langle a_{i1} \rangle}\sqrt{\langle a_{i2} \rangle}}
\, ,
\end{equation}

\begin{figure} [ht]
   \begin{center}
   \begin{tabular}{c} 
   \includegraphics[height=7cm]{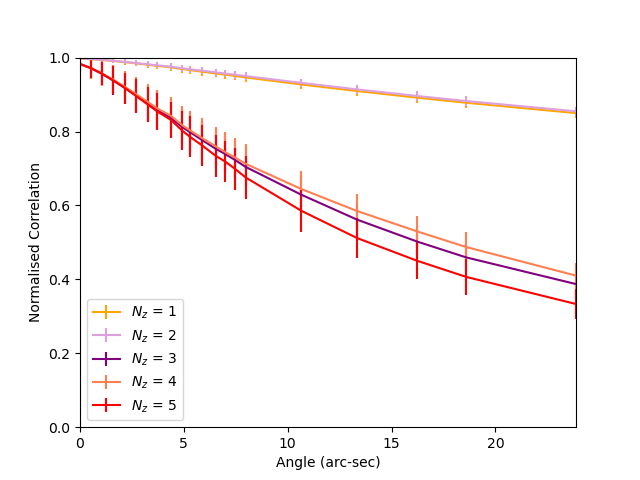}
   \end{tabular}
   \end{center}
   \caption[example] 
   { \label{fig:corr} Normalised correlation for the first 5 Zernike polynomial coefficients as a function of angle (in arcseconds). It is obtained by applying the extended PSM method on long-exposure solar images and using equation \ref{eq:corr}.}

\end{figure} 

Figure \ref{fig:corr} shows the variation in the correlation of the coefficients as a function of angle. The values are normalised with the auto-correlation values at the origin. We can see that the decorrelation increases rapidly with an increase in the order of the Zernike polynomial. This implies that lower-order terms are more correlated than higher-order terms when the two wavefronts originate from different angles on the sky.

\subsection{\texorpdfstring{Estimating the analytical form of $C_N^2(h)$ profile}{Estimating the analytical form of CN2 profile}}
\label{subsec:analytical_cn2}
Chassat \cite{1989JOpt...20...13C} established analytical expressions for the angular correlation of Zernike polynomials that we estimated in section \ref{subsec:angular_corr}. \cite{1989JOpt...20...13C} assumes a model of the turbulence strength profile to determine the expressions for the correlation given by 

\begin{equation}
\label{eq:corr_chassat}
C_{nn}(\alpha) = \left(\frac{D}{r_0}\right)^{5/3} \frac{\int_0^L dh C_N^2(h) \sigma_n(\frac{\alpha h}{R}) }{\int_0^L dhC_N^2(h)}
\, ,
\end{equation}

Equations \ref{eq:corr} and \ref{eq:corr_chassat} both describe the correlation of the coefficients. The former is for each coefficient, and the latter is the average of all coefficients of given radial order. By substituting the average correlation of the n$^\text{{th}}$ radial order we measured from equation \ref{eq:corr} in equation \ref{eq:corr_chassat}, we can get an analytical estimate for the $C_N^2(h)$ profile. If we assume a Hufnagel model \cite{Valley:80}, $C_N^2(h)$ is given by:

\begin{equation}
\label{eq:hufnagel}
C_N^2(h) = A[2.2 \text{x} 10^{-23} h^{10} e^{-h} (\frac{V_w}{\overline{V_w}})^2 + 10^{-16} e^{-h/1.5} ]\quad m^{-2/3}
\, ,
\end{equation}

\begin{figure}[H]
   \begin{center}
   \begin{tabular}{c} 
   \includegraphics[height=7cm]{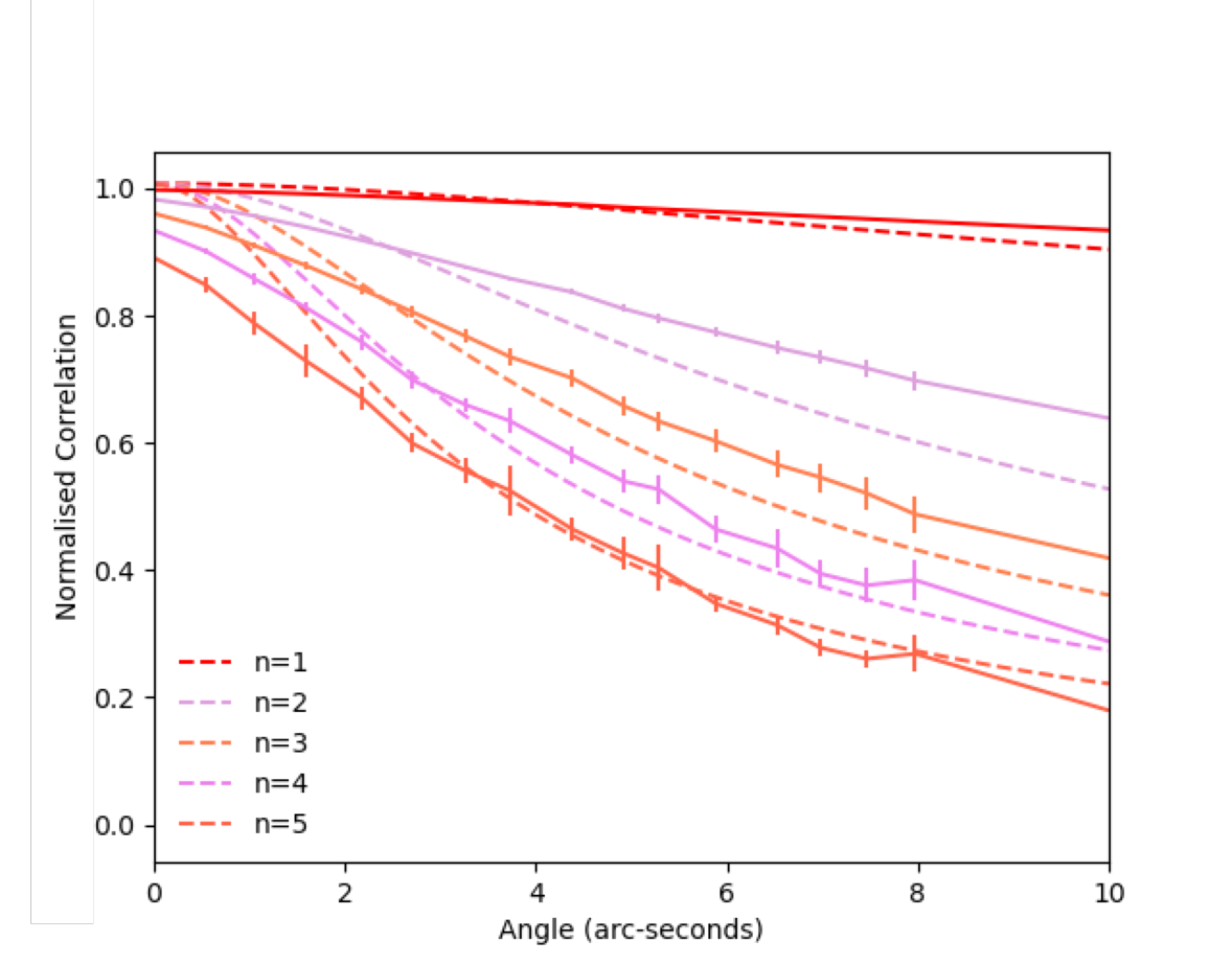}
   \end{tabular}
   \end{center}
   \caption[example] 
   { \label{fig:theoretical&measured} Normalised correlation for the first 5 radial orders of Zernike polynomial coefficients as a function of angle (in arcseconds). The dashed curves are from the theoretical expression given by Chassat \ref{eq:corr_chassat} and the solid lines are measured from extended PSM using \ref{eq:corr}.
   }
\end{figure}

Figure \ref{fig:theoretical&measured} plots the normalised correlation of coefficients for the first five radial orders of Zernike polynomials obtained using equations \ref{eq:corr} (solid curves) and \ref{eq:corr_chassat} (dashed curves). The solid lines are obtained by averaging the values of correlation for all polynomial co-efficients of a given radial order shown in Figure \ref{fig:corr}.

\begin{figure}[H]
   \begin{center}
   \begin{tabular}{c} 
   \includegraphics[height=7cm]{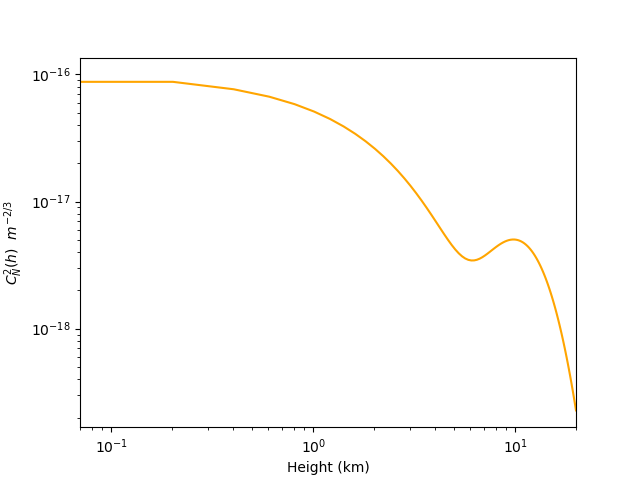}
   \end{tabular}
   \end{center}
   \caption[example] 
   { \label{fig:cn2_profile} Estimate of analytical form of $C_N^2(h)$ profile. This curve was obtained by varying the parameters used in the Hufnagel model to match/fit the theoretical and practical values of correlation.}
\end{figure}

Figure \ref{fig:cn2_profile} is the $C_N^2(h)$ profile we estimated by fitting the theoretical curves (equation \ref{eq:corr}) to the analytical expression (equation \ref{eq:corr_chassat}) by assuming a Hufnagel model.

\section{Summary and future work}
\label{sec:summary}
In this paper, we have used the extended PSM to measure the isoplanatic patch size during the daytime using long-exposure H-$\alpha$ images of the Sun and found it to be around 17 arcseconds. We also estimated the relationship between the wavefronts originating from different angular directions in terms of the correlation of Zernike polynomial coefficients. We found that the correlation decreases faster as a function of angle for higher-order polynomials. Furthermore, we have proposed a new method of constraining the $C_N^2(h)$ profile at a site using long-exposure images of the Sun. Since most Solar observatories carry out routine observations at specific wavelengths, there is a large volume of data to which this method can be applied.

We will apply the extended PSM to more data sets and estimate the isoplanatic patch size using them. We will also study the effect of the choice of model used in determining the analytical model for the $C_N^2(h)$ profile. We are also setting-up simultaneous measurements with S-DIMM+ and balloon-borne thermosonde to compare with the analytical profile we estimated from the H-$\alpha$ images.

\printbibliography 
\end{document}